\documentclass[epj]{svjour}
\usepackage{graphicx}
\usepackage{amssymb}
\usepackage{bm}
\begin{document}

	\title{Design of artificial genetic regulatory networks with multiple delayed adaptive responses}

	\author{Pablo Kaluza\inst{1,}\inst{2} \and Masayo Inoue\inst{3,}\inst{4}}

	\institute{National Scientific and Technical Research Council \& Faculty of Exact and Natural Sciences, National University of Cuyo, Padre Contreras 1300, 5500 Mendoza, Argentina. \and Abteilung Physikalische Chemie, Fritz-Haber-Institut der Max-Planck-Gesellschaft, Faradayweg 4-6, 14195 Berlin, Germany. \and  Molecular Profiling Research Center for Drug Discovery, National Institute of Advanced Industrial Science and Technology, 2-4-7 Aomi, Tokyo 135-0064, Japan. \and Cybermedia Center, Osaka University, Toyonaka, Osaka 560-0043, Japan.}

	\mail{pkaluza@mendoza-conicet.gob.ar}

	\date{Received: date / Revised version: date}

	\abstract{
		Genetic regulatory networks with adaptive responses are widely studied in biology. Usually, models consisting only of a few nodes have been considered. They present one input receptor for activation and one output node where the adaptive response is computed. In this work, we design genetic regulatory networks with many receptors and many output nodes able to produce delayed adaptive responses. This design is performed by using an evolutionary algorithm of mutations and selections that minimizes an error function defined by the adaptive response in signal shapes. We present several examples of network constructions with a predefined required set of adaptive delayed responses. We show that an output node can have different kinds of responses as a function of the activated receptor. Additionally, complex network structures are presented since processing nodes can be involved in several input-output pathways. 
	}

	\authorrunning{P. Kaluza}
	\titlerunning{Networks with multiple adaptive responses}
	\maketitle

	\section{Introduction}
	
	Gene regulatory networks of living organisms can present a particular kind of response, adaptive response, against environmental changes. This response is important in order to retain the operation and functionality of the biological systems.  The expression levels of some genes inside the cell show changes as a response to an external stimulus. These changes return later to pre-stimulus values presenting adaptation to the new environmental conditions. 
	The change on the gene expression level can be an up-regulation (increment) or a down-regulation (decrement) as a function of  the type of stimulus \cite{Gasch1,Causton,Gasch2}. And in occasions, these adaptive responses can appear delayed with respect to the activation by the external signal \cite{Nishikawa}.
	
	There have been many theoretical studies for adaptive responses with simple models composed with a few elements \cite{Siggia,Koshland,Knox,Inoue_2011}. Especially, detailed analysis of small genetic networks of three nodes showing adaptive responses has been performed \cite{Trusina,book_zhang}. 
	The identification of these systems or motifs with adaptive responses is possible by an exhaustive searching of all possible combinations of pattern connections, however, when the number of nodes is not small, this combinatorial analysis cannot be performed in a rational time. 
	In this case, stochastic algorithms of optimizations as genetic ones have been used to construct relatively large networks with adaptive responses \cite{Siggia,Inoue}. Moreover, in both small and large network cases, an adaptive response between a specific pair of input-output nodes has been studied, i.e. the input signal is applied to only one node (receptor) and the adaptive response is computed only for one output node.
	
	In this work we propose to construct networks with adaptive responses between several input receptors and several output nodes. We even study adaptive responses with time delays; the output nodes show no response at all for a period of time after stimulus to input receptors, but then they suddenly start to show adaptive responses. Such responses with time-delay are observed ubiquitously \cite{Nishikawa,delayed_response}, however, adaptive responses with time delays have been rarely studied. 
	
	We employ relatively large networks with several tens of nodes. They are not so much large, but enough large to make it difficult to study a full search as in three nodes cases. The networks with delayed adaptive responses imposed complex architectures that cannot be designed by trial and error of all possible combinations neither by a rational design because of its number of nodes. For example, a network with three types of $20$ possible connections has approximately $3 ^ {20} \approx 3.48 \times 10^{9}$ configurations. 
	Therefore, we propose to use a version of the Metropolis algorithm. In a given network with some actual output we measure its error with respect to a target set of signals, and we try to reduce that error. This kind of optimization, Metropolis-like methods, have been used in the construction of genetic networks \cite{yanagita_PRE_2012,kobayashi_PRE_2011,yanagita_PRE_2010,kobayashi_EPJB}, and, flow processing networks \cite{kaluza_EPJB,kaluza_CHAOS,kaluza_pre,kaluza_EPL}.

	We show in this article that it is possible to construct networks with delayed adaptive responses between multiple input receptors and multiple output nodes.
	First, we studied adaptive responses without delay but between multiple input receptors and multiple output nodes. We show that an output node can present different types of responses according to the input receptor activated, although the routes connecting each input receptor and the output node are often overlapped. Next, we study delayed adaptive responses between multiple input and output nodes. We constructed networks not only with different delay times for an output node according to different input receptors but also with ones among different output nodes with respect to an input node. Thus, we could construct networks with any delayed adaptive responses using our algorithm. 
	
	The paper is organized as follows: in section \ref{sec_model} we present the network model, the dynamics of the nodes, the cost function or error of a network and the annealing method used in the optimizations. In section \ref{sec_numerical} we show several examples of construction of networks with different target adaptive responses and network sizes. Finally, in section \ref{sec_conclusion} we present the final discussions and results.

	\section{Models and methods}
	\label{sec_model}
	
\begin{figure}[!ht] 
		\begin{center}
			\includegraphics[width=1.0\columnwidth, clip]{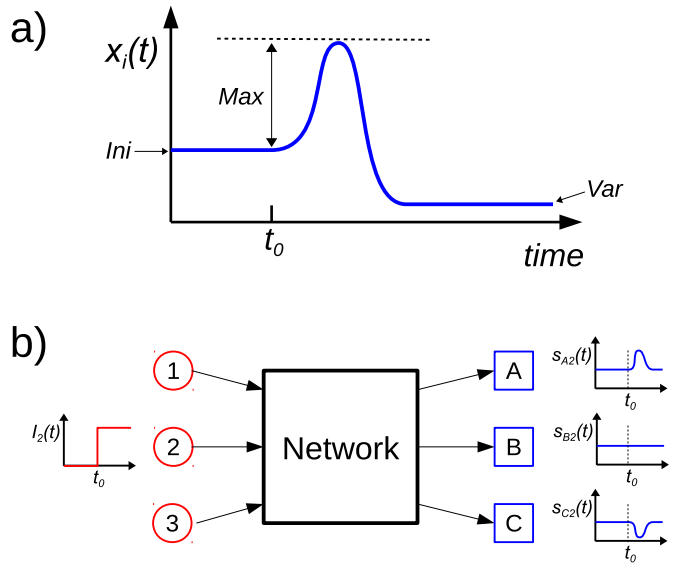}
			\caption{a) Example of an adaptive signal $x_i(t)$. At time $t_0$ the external signal $I(t) \ne 0$ is applied and the node $i$ starts its response. The activity grows a quantity $Max$ from its initial value $Ini$ and evolves to an value $Var$. b) Representation of our problem. A network with three input receptors (circular nodes $1$, $2$ and $3$) and three output nodes (squares $A$, $B$ and $C$). Input signal $I_2(t)$ activates the network at time $t_0$ and output signals $s_{A2}(t)$, $s_{B2}(t)$ and $s_{C2}(t)$ are expressed on the output nodes.}
			\label{fig_fig01} 
		\end{center}
	\end{figure}

	The adaptive response of a gene in a regulatory system is a process where the gene $i$ changes its level of expression $x_i(t)$ when some external signal $I_k(t)$ activates at $t=t_0$ certain receptors of the network. This change in the expression level can be positive (up-regulation) or negative (down-regulation) and it has a pulse-like shape as a function of time, i.e., the expression level returns close to the pre-activation value before the external signal was applied. Note that the external signal keeps activating the receptors for any $t>t_0$. The adaptive response can start delayed with respect to the activation of the receptor by the external signal at $t_0$. Figure \ref{fig_fig01}a shows a schematic picture of an adaptive response. 
	
	An adaptive response can be characterized by its shape with introducing these three values, $Ini$, $Max$, and $Var$ \cite{Siggia}. $Ini$ is the steady state value of $x_i(t)$ before the application of an external stimulus. $Max$ is the maximal absolute change from the $Ini$ value and $Var$ is the new steady state value after the application of the stimulus. It is clear from Fig. \ref{fig_fig01}a that an adaptive response is well defined with larger $Max$ and smaller $|Ini - Var|$. 
	The response starts immediately after the application of the input signal at $t=t_0$ in Fig. \ref{fig_fig01}a, but it can start delayed as we consider in this paper. 

	In this work the aim is to design networks with several input receptors and several output nodes presenting adaptive responses. Figure \ref{fig_fig01}b shows an example of these systems. A network $G$ has $N$ nodes with $N_{in}$ input receptors, $M$ middle nodes and $N_{out}$ output nodes. These networks process input signals $I_k(t)$ ($k=1,..., N_{in}$) and generate responses $s_{jk}(t)$ ($j=1,...,N_{out}$, $k=1,...,N_{in}$) on the output nodes. We consider the input signals acting only one at the time, thus, the response matrix $\mathbb{R} = \{s_{jk}(t)\}$ describes the network response for these input signals by the network $G$. Since all the nodes follow the same dynamics and the initial conditions of the dynamical system are fixed, the matrix response is a function of the network structure, i.e. the pattern of connections.
		
	Generally, with a random connection matrix, output nodes do not show adaptive responses and often show monotonic evolutions to fixed points or oscillations. In addition, response with time delay is hardly realized.

	\subsection{Network model}
	The network model we consider is essentially the one used in \cite{Inoue}. However, we extent that previous model in order to have a layered-like network structure with several input receptors and several output nodes (see Fig. \ref{fig_fig01}b). Following, we present in detail the technical aspect of the model. The biological interpretation and argumentation of the validity of this model can be found in the previous reference.

	\subsubsection{Network structure}
\begin{figure}[!ht] 
		\begin{center}
			\includegraphics[width=1.0\columnwidth, clip]{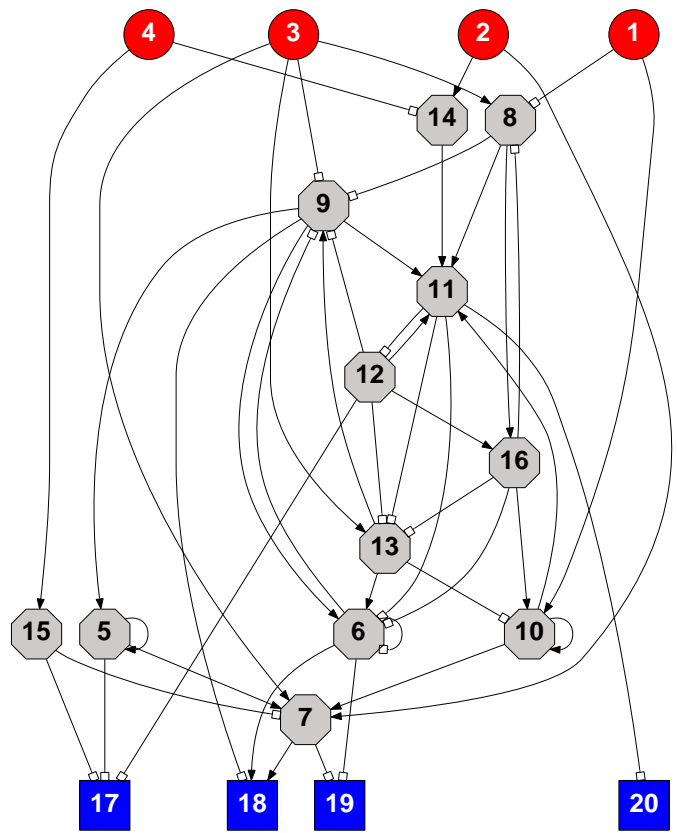}
			\caption{Example of network. There are $N_{in}=4$ receptors (red circular nodes $1$, $2$, $3$ and $4$), $M=12$  middle nodes (gray octagons), and $N_{out}=4$ output nodes (blue square nodes $17$, $18$, $19$ and $20$). Connections ending in filled arrows are excitatory and connections ending on white empty squares are inhibitory.} 
			\label{fig_fig02} 
		\end{center}
	\end{figure}

	We use a regulatory network model composed with nodes interacting each other. In a network $G$, there exist $N$ nodes in total and the nodes are classified into three types; $N_{in}$ input nodes (receptors) receiving the external stimulus, $N_{out}$ output nodes showing the final responses against the stimulus, and $M$ middle nodes processing the stimulus from receptors to output nodes. Figure \ref{fig_fig02} shows an example of network.
	
	Nodes are connected with the following rules. Input nodes can be connected only with middle nodes (from input nodes to middle nodes). Middle nodes can be connected with middle nodes (from middle nodes to middle nodes) and with output nodes (from middle nodes to output nodes). No other types of connections are permitted and only middle nodes can have self-connections. Therefore, output nodes have only incoming connections from middle nodes.
	
	Only one directed link can exist between two nodes and each connection can be excitatory or inhibitory. To describe the network architecture, we use a connection matrix $\mathbb{C}$; the element $C_{ij}$ represents an interaction from node $j$ to node $i$. $C_{ij}$ takes $1$, $-1$, or $0$ depending on whether the connection is excitatory, inhibitory, or non-existent.
	
	\subsubsection{Network dynamics}
	\label{subsubsec_net_dyn}
	
	A node $i$ has an internal variable $x_i(t)$ for its response level with a time evolution given by 
	
	\begin{equation}
		\frac{dx_i}{dt} = \frac{1}{1 + \exp (-\beta y_i ) } - \gamma x_i + \alpha.
		\label{equ_model_x}
	\end{equation}
	
	The first term represents interactions with other nodes and the second term represents degradation, while $\alpha$ is a small output representing spontaneous response. $y_i$ shows the total input signal to node $i$ $(i=1,...,N)$ and is given by
	
	\begin{equation}
		y_i = I_k(t) \delta_{ik} + \sum_{j=1}^{N} C_{ij}x_{j},
		\label{equ_model_y}
	\end{equation}
	
	\noindent
	with $\delta_{ik}=1$ (for $i=k$), 0 (for $i \neq k$) and $k=1,\cdots , N_{in}$. Thus, the external stimulus $I_k(t)$ is applied only to the input nodes.
	
	We set the following parameter values: $\beta =10$, $\gamma=1$ and $\alpha = 0.01$. 
	The external stimulus $I_k(t)= 0$ for $t<t_0$ and $I_k(t)= I ^{\ast} $ for $t \geqslant t_0$. $I ^{\ast}$ need to be enough large to activate input nodes and we set $I ^{\ast} = 5$ in this work. The value $t_0$ indicates the instant when the input node $k$ is activated. We use same time evolution (eq.(\ref{equ_model_x})) and same parameters for all nodes regardless of the type of nodes. 
	We thus fix the parameter values concerning to the dynamics of nodes, while we change the number of nodes ($N$, $N_{in}$, $M$, and $N_{out}$) in each case and study evolution of the connection matrix $\mathbb{C}$.
	
	All nodes are put at $x_i(0)=0.5$ as initial conditions and evolve to a steady state or an oscillatory regimen according to eq.(\ref{equ_model_x}) without external stimulus under a connection matrix $\mathbb{C}$. We use this initial condition $\{\bm{x_0}\}$ for all cases in this paper.
		
	The first term in eq.(\ref{equ_model_x}) changes from $0$ through $1$ according to $y_i$. For full inhibitory interaction ($y \ll -1$), it approaches to $0$ and therefore $x_i(t) \to \alpha/\gamma$. On the other hand, for full excitatory interaction ($y \gg 1$), it approaches to $1$ and $x(t) \to (1+\alpha)/\gamma$. As a result, $x_i(t)$ varies between these two values, $x_i(t) \in [0.01, 1.01]$ as $\alpha=0.01$ and $\gamma=1$. In addition, when there is no interaction ($y=0$), $x(t) \to (0.5+\alpha)/\gamma = 0.51$. 
	
	We have to note now that an output signal $s_{jk}(t)$ ($j=1,...,N_{out}$ and $k=1,...,N_{in}$) corresponds to the variable $x_i(t)$ of the output node $i$, that is, $s_{jk}(t) \equiv x_i(t)$ for $i= N_{in} + M + j$, when the input node $k$ is activated. 
	
	\subsection{Error function}
	
	Our task is to generate networks with a specific set of output signals. Thus, we need to define some kind of distance between the actual output response of a given network and the target response we desire to construct. We call the set of target signals  $\mathbb{T}$ (target pattern). On the other hand, a given network $G$ with structure $\mathbb{C}$ presents an actual set of output signals $\mathbb{R}=\{s_{ij}(t)\}$ (response). The distance $\epsilon$ between the target pattern and the actual response is defined as the error of the network $G$ with respect to the target pattern, i.e, $\epsilon(G) = | \mathbb{T} - \mathbb{R}|$. Matrices $\mathbb{T}$ and $\mathbb{R}$ have elements as temporal signals. Thus, in order to compute the distance between their elements we measure how different are the actual output signals with respect to target ones. In order to perform this calculation, we proceed as follows.
	
\begin{figure}[!ht] 
		\begin{center}
			\includegraphics[width=1.0\columnwidth, clip]{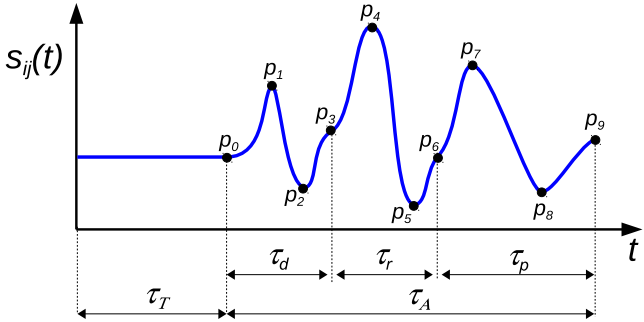}
			\caption{Schematic representation of an output signal $s_{ij}(t)$. We show the time intervals and their characteristic points $p$'s. The time intervals are: transient $\tau_T$ and adaptive response $\tau_A$, and, the latter one is divided into delay $\tau_d$, response $\tau_r$ and post-response $\tau_p$ subintervals. }  
			\label{fig_fig03} 
		\end{center}
	\end{figure}

	We split the output response $s_{ij}(t)$ into several temporal intervals and evaluate each of them (Fig. \ref{fig_fig03}). During the transient interval $\tau_T$ with $t<t_0$, the external stimulus $I_j(t)=0$ and $s_{ij}(t)$ is stabilized, ideally, on a stable fixed point. At $t=t_0$ the stimulus is applied, and $s_{ij}(t)$ starts to show some response. We call this interval with expected adaptive response as $\tau_A$ ($t \geqslant t_0$) and divide into three subintervals: a delay interval $\tau_d$, a response interval $\tau_r$, and a post-pulse interval $\tau_p$. We expect the adaptive response is realized during the response interval $\tau_r$ and $s_{ij}(t)$ stays almost constant 
	during $\tau_d$ and $\tau_p$.

	For each of these subintervals, we define the initial, the final, the maximum and the minimum values as shown in Fig. \ref{fig_fig03} with the points $p$'s. We now define error function for realizing a target output response ($\mathbb{T}$) with these values. We first define the error function for each subinterval and then we combine them into one final quantity. 
	
	During the delay interval $\tau_d$, the output response must be constant, thus, we define the error function for this interval as the difference between the maximum $p_1$ and minimum $p_2$ values:
	\begin{equation}
		\epsilon^d = |p_1-p_2|.
		\label{equ_error_delay}
	\end{equation}
	
	\noindent
	We find a similar situation for the post-pulse interval $\tau_p$; the output signal must be constant and the error function for this interval is defined as the difference between the maximum $p_7$ and minimum $p_8$ values:
	\begin{equation}
		\epsilon^p = |p_7-p_8|.
		\label{equ_error_postresponse}
	\end{equation}
	
	For the response interval $\tau_r$, we set the three situations as already explained: a constant response, a positive adaptive response (up-regulation), and a negative adaptive response (down-regulation). We set different error functions for each case. 
	
	In case of a constant response, the error function is given by the difference between the maximum $p_4$ and minimum $p_5$ values:
	\begin{equation}
		\epsilon^r = |p_4-p_5|.
		\label{equ_error_response_c}
	\end{equation}
	
	\noindent
	In case of either adaptive responses, considering the characterization with $Ini$, $Max$, and $Var$ in Fig. \ref{fig_fig01}a, we define the error function as follows. $Ini$ ccorresponds to $p_3$ and $Var$ to $p_6$. As for $Max$, it corresponds to $p_4 - p_3$ in positive adaptive case and to $p_3 - p_5$ in negative case. Then, in case of a positive adaptive response (up-regulation), we define the error function as
	\begin{equation}
		\epsilon^r = 1.0 - \big\{ P + (1- |p_3 - p_6|) + \{1- (p_3 - p_5)\}  \}/3.0.
		\label{equ_error_response_p}
	\end{equation}
	with
	\begin{equation}
		P = \left\{ 
		\begin{array}{l l}
		2(p_4 - p_3) & \mbox{ if } (p_4 - p_3) \le 0.5\\
		2 (1- (p_4 - p_3)) & \mbox{ otherwise}\\
		\end{array} \right.
	\end{equation}
	
	\noindent
	Similarily, in case of a negative adaptive response (down-regulation), we define as
	\begin{equation}
		\epsilon^r = 1.0 - \big\{ Q +(1- |p_3 - p_6|) + \{1 - (p_4 - p_3)\} \}/3.0;
		\label{equ_error_response_n}
	\end{equation}
	with
	\begin{equation}
		Q = \left\{ 
		\begin{array}{l l}
		2(p_3 - p_5) & \mbox{ if } (p_3 - p_5) \le 0.5\\
		2 (1- (p_3 - p_5)) & \mbox{otherwise}\\
		\end{array} \right.
	\end{equation}

	With these definitions we impose the condition that the minimum error is reached for adaptive responses with a pulse amplitude $Max=0.5$ (see Fig. \ref{fig_fig01}). We impose this restriction since the signals $s_{ij}(t) \in [0.01, 1.01]$ (see sec. \ref{subsubsec_net_dyn}) and an output node may need to show both positive and negative adaptive responses as a function of the stimulated receptors. 
	
	The total error $\epsilon_{ij}$ of an output response $s_{ij}(t)$ is defined as
	\begin{equation}
		\epsilon_{ij} = a_d \epsilon^d_{ij} + a_r \epsilon^r_{ij} + a_p \epsilon^p_{ij}.
		\label{equ_error_i}
	\end{equation}
	
	\noindent
	The coefficients $a_d$, $a_r$, $a_p$ need to satisfy $a_d + a_r + a_p = 1$ according to relative importance upon the total error. We set $a_d = 1/10$, $a_r = 8/10$, and $a_p = 1/10$ throughout the paper. With this definition, the error during the response interval ($\tau_r$) has more importance than the errors during the other two intervals. We set this election because the main problem is to generate the pulse in the response interval. The other two errors are mainly added in order to avoid oscillatory responses. Finally, as we have $N_{in}$ input receptors and $N_{out}$ output nodes, the total error of a network $G$ is given by
	
	\begin{equation}
		\epsilon(G) = \frac{1}{N_{in} N_{out} } \sum_{i=1}^{N_{in}} \sum_{j=1}^{N_{out}} \epsilon_{ji}.
		\label{equ_error_G}
	\end{equation}

	\subsection{Optimization construction}

	The process to construct a network $G$ with a predefined response $\mathbb{T}$ seems just an optimization problem where we need to find a minimum of the error function $\epsilon(G)$. This optimization can be performed by several different techniques. In our case, we employ an annealing algorithm \cite{annealing}.
	
	The algorithm consists of the following steps:
	
	\begin{enumerate}
	\item Take a network $G$ with error $\epsilon$.
	\item Apply an evolutionary mutation to $G$, obtaining $G'$ with error $\epsilon'$.
	\item Calculate $\Delta \epsilon = \epsilon'- \epsilon$.
	\item If $\Delta \epsilon \le 0$ accept the mutation making $G = G'$. If $\Delta \epsilon > 0$ accept the mutation with a probability $\exp(-(\Delta \epsilon)/(\sigma \epsilon))$
	\item Return to step 1.
	\end{enumerate}

	\begin{figure}[!ht] 
	\begin{center}
		\includegraphics[width=1.0\columnwidth, clip]{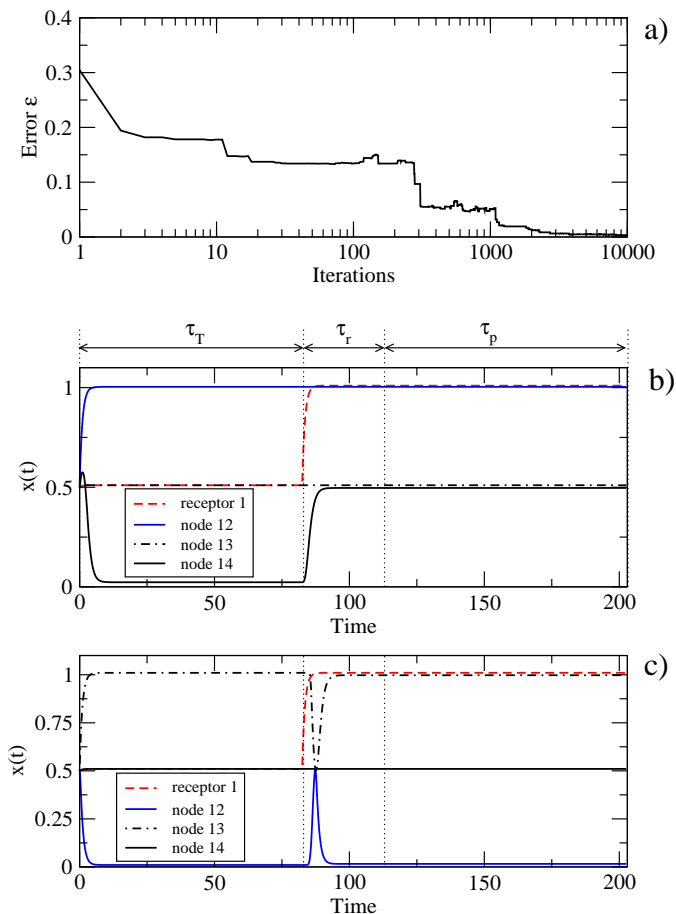}
		\caption{Examples of network evolution. a) Error as a function of the number of iterations.  b) Output signals of the nodes of the initial network (see fig \ref{fig_fig05}a). c) Output signals of the nodes of the final network (see fig \ref{fig_fig05}b). }
		\label{fig_fig04}
	\end{center}
	\end{figure}

	\noindent
	This process is repeated during a fixed number of iterations or until we find an error smaller than some given threshold value.  In this algorithm, $\sigma \epsilon$ plays the role of temperature and decreases with the error approaching to zero. The parameter $\sigma$ controls the importance of temperature and an optimal value for the convergence exists in general. In each optimization trial, we start with a random network connected with a probability of $p=0.1$. A link can be excitatory or inhibitory with the same probability.
	
	\begin{figure}[!ht] 
	\begin{center}
		\includegraphics[width=0.9\columnwidth, clip]{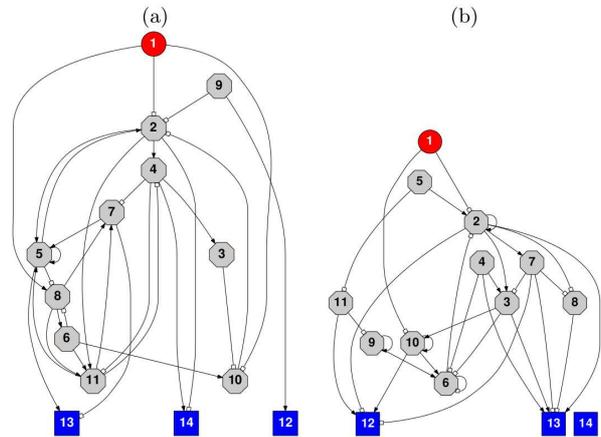}
		\caption{Initial (a) and final (b) networks from the example shown in Fig. \ref{fig_fig04}.}
		\label{fig_fig05}
	\end{center}
	\end{figure}
	
	Figures \ref{fig_fig04} and \ref{fig_fig05} show an example of this process of network construction for a system with $N_{in}=1$, $M=10$ and $N_{out}=3$ nodes. The target output response we consider is 
	
	\begin{equation}
		\mathbb{T} = \left[
		\begin{array}{c}
			A_{0} \\
			R_{0} \\
			0
		\end{array}
		\right].
		\label{equ_target_example}
	\end{equation}
	
	\noindent
	The elements of the target matrix $\mathbb{T}$ are given with the notation $X_{\tau_d}$ with $X=A, R, 0$ representing an adaptive response with up-regulation, an adaptive response with down-regulation and a constant response respectively. The subindex represents the delay interval $\tau_d$. Note that the column in the target matrix has the input node index $i=1,...,N_{in}$, and the rows show the output node indexes $j=1,...,N_{out}$. Therefore, eq.(\ref{equ_target_example}) indicates that the first output node (node-$12$ in Fig.\ref{fig_fig05}) shows an up-regulation with $\tau_d=0$, the second one (node-$13$ in Fig.\ref{fig_fig05}) shows a down-regulation with $\tau_d=0$, and the last one (node-$14$ in Fig.\ref{fig_fig05}) shows a constant signal. 
	
	Figure \ref{fig_fig04}a presents the error $\epsilon$ as a function of the number of iterations. Figure \ref{fig_fig04}b presents the output responses $\mathbb{R} = \{s_{ij}(t)\}$ for the initial random network shown in Fig. \ref{fig_fig05}a. We can observe that they are far from adaptive responses. Figure \ref{fig_fig04}c shows the output responses of the final network shown in Fig. \ref{fig_fig05}b and the target pattern is realized. The vertical dashed lines indicates the beginning and the end of the response interval $\tau_r=30$.

	\subsubsection{Evolutionary mutation}
	
	We consider two different schemes of mutations for the optimization of the networks. The first one is called \textit{link mutation} and it consists of adding a new link or removing an existing one with equal probability in each iteration. A new link can be excitatory or inhibitory with equal probability. This scheme has been successfully used in our previous work of networks with time-programmed responses \cite{kaluza_EPJB}. The main characteristic point of this scheme is that the total number of links cannot be controlled during the optimization process. 
	
	In order to control the total number of links, we use the second scheme called \textit{rewiring mutation}. This mutation consists of rewiring of the links. That is, we delete a link randomly and we create a new one between two randomly chosen nodes without existing connection between them. As for the type of the new link, excitatory or inhibitory, we do not keep the previous type and choose randomly with equal probability. As a result, the total number of links is fixed, but the number of excitatory and inhibitory connections are not preserved during the optimization course.
	
	The number of nodes is conserved in both schemes. It can happen that some middle nodes have only input connections at the end of the optimization. These nodes can be removed from the network without changing the responses of the output nodes.

	\subsubsection{Numerical integration and evaluation of the error}

	In order to find the error of a given network with respect to the target, it is necessary to integrate the system of differential equations (\ref{equ_model_x}) in each iteration of the evolutionary algorithm of optimization. Although, it is a normal procedure, we need to consider several points given the particular characteristics of this system.

	The system we study shows oscillations quite frequently. This situation cause a problem that oscillatory responses can be computed like adaptive responses if the time intervals for computing the error function are shorter than their periods. To avoid this, we take $\tau_p = \tau_d + 3\tau_r$, thus, $\tau_A = 2(\tau_d + 2 \tau_r)$ (see Fig. \ref{fig_fig03}). These conditions ensure the second pulse occurs before the end of $\tau_p$ in case of periodic responses. A second pulse set the error $\epsilon^p$ to non-zero and increase the total error. Therefore, oscillatory responses are eliminated from the final networks. 
	
	It is possible to generate adaptive response with long delays after the input signal is applied. This indicates that the dynamics can be quite slow under some conditions. Therefore, we need to set enough longer transient interval for the system to relax to the steady state. Then, we set $\tau_T = \tau_d + 2\tau_r$. This interval is consistent with the requited delay of the response.  
	
	We find that fixing the initial condition and the time intervals for the signal, it is possible to generate a pulse (adaptive response) as we require without need the external input signal $I_k(t)$. A simple way to avoid this situation is taking the transient interval $\tau_T$ different in each new integration. We used an effective transient interval for the integrations by taking the time interval $\tau_T$ plus a random extra time chosen between zero and $\tau_r$. 
	
	The construction of a network by this method is demanding from the computational point of view since we need to integrate the system in each iteration of the optimization. On the other hand, Eq. (\ref{equ_model_x}) presents a smooth dynamics without strong changes. For these reasons we employ an Euler algorithm to integrate the system during the optimization. We use $\Delta t = 0.01$. In order to validate the results, at the end of the simulation we evaluate again the final network dynamics with a Runge-Kutta method of fourth order with $\Delta t = 0.001$. In general, we do not find any significant change on the results and both methods give the same time evolution.

	\section{Numerical results}
	\label{sec_numerical}
	
	We present in this section several examples of network constructions for systems with different sizes and delayed responses. Since our main goal is to show the effectives of the algorithm of optimization, we do not analyze in detail the network properties of the constructed systems, and we focus on the interesting example that can be constructed. The obtained networks are collected in the supplementary data file \textit{networks\_adaptive\_response.nets} by their adjacency matrices and target required patterns. The objective of this network collection is to provide easy access to the constructed system for their evaluation.

	\subsection{Adaptive responses without delays}
	
	In our first set of examples we consider adaptive responses without delays ($\tau_d=0$), and we study influences of network sizes. 
	
	\subsubsection{Small networks with one receptor and one output node}
	\label{subsub_section_nodelay_small}
	
	We consider networks with one receptor, three middle nodes and one output node. These networks are the smallest ones for which we can find adaptive responses. Our minimum networks have $5$ nodes in total and they are larger than the smallest adaptive network reported in \cite{Trusina}. This difference on the number of nodes is basically due to the simple dynamics of each element with fixed parameters (eq. (\ref{equ_model_x})) and the layered structure of our model that imposes restrictions on the connection pattern. The target signal we consider do not have a delay ($\mathbb{T} = A_0$) and the response interval has a time windows $\tau_r=30$ where an adaptive positive pulse must hold. During the optimization we use the link mutation scheme, thus, the number of connections can vary. The total number of iterations is $1 \times 10^4$.
	
\begin{figure}[!ht] 
		\begin{center}
			\includegraphics[width=1.0\columnwidth, clip]{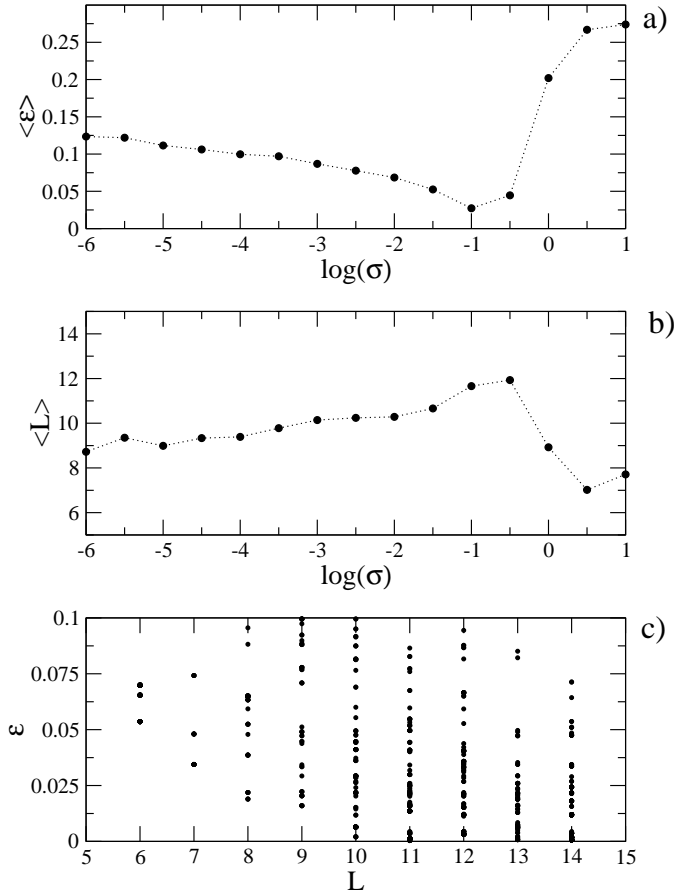}
			\caption{a) Mean error $\langle \epsilon \rangle$ as a function of the temperature parameter $\sigma$. b) Mean number of links $\langle L \rangle$ as a function of the temperature parameter $\sigma$. c) Errors $\langle \epsilon \rangle$ as a function of the number of links.}
			\label{fig_fig06} 
		\end{center}
	\end{figure}

	We run several realizations in order to create ensembles of $200$ networks each by using different values of the temperature parameter $\sigma$. In each realization a random initial network is considered with a connectivity $p=0.1$. Figure \ref{fig_fig06}a presents the mean error $\langle \epsilon \rangle$ of each ensemble as a function of $\sigma$. We observe that there is a minimum for $\log( \sigma) \approx -1$. This value corresponds to the optimum temperature parameter, for larger values the optimizations do not converge, whereas smaller values stack the systems in local minima where the solutions are not the best ones.
	
	\begin{figure}[!ht] 
		\begin{center}
			\includegraphics[width=1.0\columnwidth, clip]{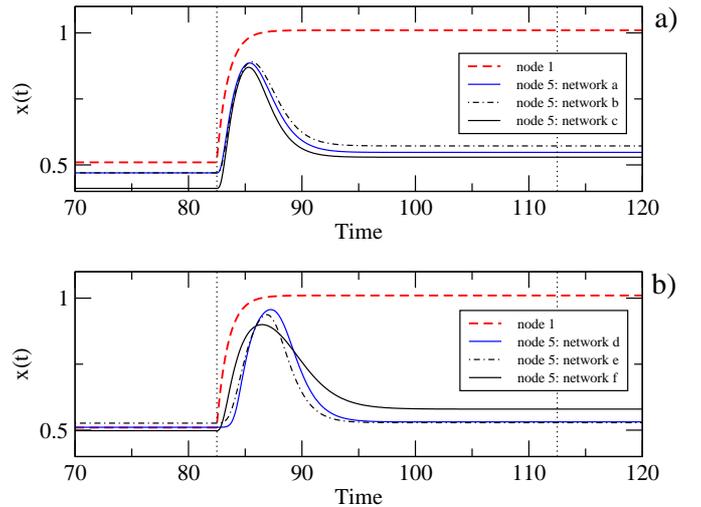}
			\caption{Output signals of networks from Fig. \ref{fig_fig08}. Vertical dashed lines indicate the time windows $\tau_r=30$.}
			\label{fig_fig07} 
		\end{center}
	\end{figure}
	
	We find several networks able to perform adaptive responses. Since the number of links during the evolution can vary due to the mutation scheme, the successful final networks have in general different number of links. Figure \ref{fig_fig06}b presents the mean number of links $\langle L \rangle$ as a function of the temperature parameter $\sigma$. We observe that networks with the higher number of links are located in $\log(\sigma) \approx -1$, where the mean error $\langle \epsilon \rangle$ is the minimum. 
	
	Figure \ref{fig_fig06}c presents the errors $\epsilon$'s as a function of the number of links for all the network constructions with small error. We observe that the solutions with smaller errors are located for larger number of links. This result means that in general it is easier to generate an adaptive response with dense networks than with sparse ones. Note that the highest number of links for these networks is $15$.

	\begin{figure}[!ht] 
	\begin{center}
		\includegraphics[width=1.0\columnwidth, clip]{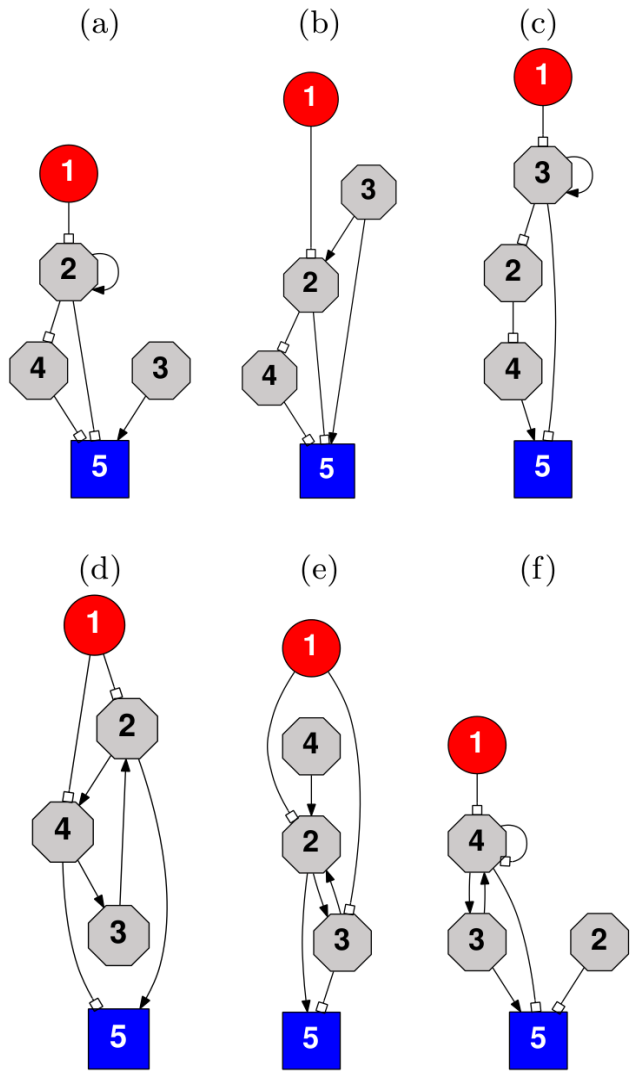}
		\caption{Networks with adaptive responses and few links. The errors of these networks are: $\epsilon_a = 0.065$, $\epsilon_b = 0.070$, $\epsilon_c = 0.054$, $\epsilon_d = 0.034$, $\epsilon_e = 0.048$ and $\epsilon_f = 0.074$.}
		\label{fig_fig08}
	\end{center}
	\end{figure}

	We are particularly interested in solutions with few links since they are more realistic and can be considered for implementations from the point of view of synthetic biology. Examples of output signals are shown in Fig. \ref{fig_fig07} for networks with six (a, b and c) and seven links (d, e and f). Networks are shown in Fig. \ref{fig_fig08}. In all the cases the adaptive pulse start immediately after the activation of the receptor. The expression level of the receptor grows rapidly to the maximum level. Note that the error is not zero since the pulses have amplitudes smaller than $0.5$, and because the signals after the pulses do not return to the pre-stimulus values exactly.

	\subsubsection{Networks with three input receptors and three output nodes}
	\label{subsub_section_3x3_1}
	
	For our second example of network constructions we consider systems with $N_{in}=3$ receptors, $M=15$ middle nodes and $N_{out}=3$ output nodes. The required target pattern is shown in the matrix \ref{matrix_activation}.

	\begin{equation}
		\mathbb {T} = \left[
		\begin{array}{c c c}
			R_0 & 0 & A_0 \\
			0 & A_0 & R_0 \\
			A_0 & R_0 & 0
		\end{array}
		\right].
		\label{matrix_activation}
	\end{equation}
	
	\noindent
	In this target pattern each output node must produce three different responses depending on the receptor which is activated. The responses (pulse) must be located in a time windows of $\tau_r = 30$ after the onset of the external signals $I_k(t)$. 
	
\begin{figure}[!ht] 
		\begin{center}
			\includegraphics[width=1.0\columnwidth, clip]{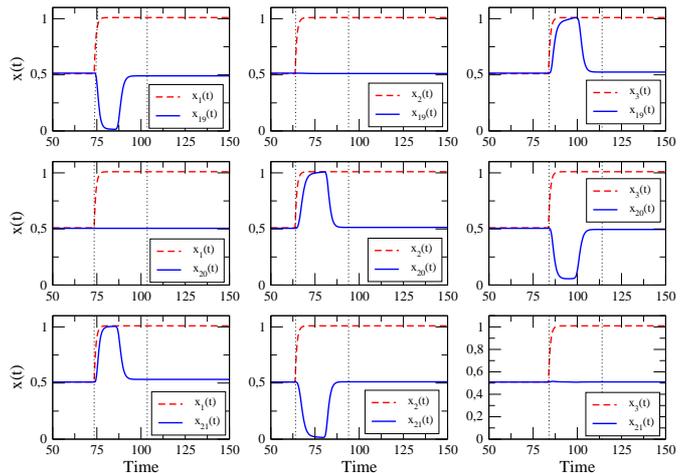}
			\caption{Output signals (solid blue curves) of the network from Fig. \ref{fig_fig10} with a target pattern shown in matrix \ref{matrix_activation}. The red dotted curves present the expression of the activated receptor.  The response windows $\tau_r=30$ is shown between vertical dashed lines.}
			\label{fig_fig09} 
		\end{center}
	\end{figure}

\begin{figure}[!ht] 
		\begin{center}
			\includegraphics[width=1.0\columnwidth, clip]{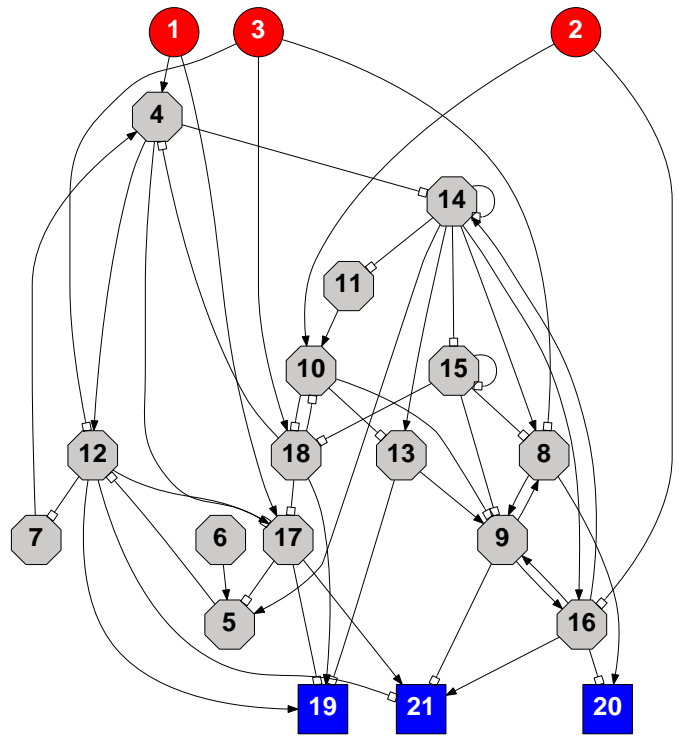}
			\caption{Designed network with the target responses shown in matrix \ref{matrix_activation} and output signals shown in Fig. \ref{fig_fig09}. The network error is $\epsilon=0.01$.}
			\label{fig_fig10} 
		\end{center}
	\end{figure}

	We used the rewiring mutation with $50$ connections. Similarly to the previous example we have run several realization with different values of $\sigma$ and we find that the best convergence is for $\log(\sigma) \approx -2$. Figure \ref{fig_fig09} presents the responses $\{s_{ij}(t)\}$ of the constructed network shown in Fig. \ref{fig_fig10}. We observe that the required target pattern can be well reproduced. We present in red dotted lines the expression of the activated receptor in order to show the onset of the external activation.
	
	The evolution find a solution where the responses of all the output nodes after the transient interval stay close to $x_i(t) \approx 0.51$, allowing to maximize both pulse amplitudes, the up-regulations (activations) and the down-regulations (repressions). We have performed a similar study by employing the link mutation scheme where the number of connections during the evolution is not fixed. In this case, we also found solutions with small error but the number of connections were larger.

	\subsection{Adaptive responses with delays}
	
	The second set of examples considers constructions of networks with delayed responses ($\tau_d \ne 0$) and different network sizes.

	\subsubsection{Small networks with one receptor and one output node}
	\label{sec_1_1_delays}
	
	We consider small networks with one receptor and one output node with delayed responses. We construct networks with $M=5$ middle nodes and $L=15$ links. In the optimization we use $\log( \sigma ) = -2$, and, we apply the rewiring mutation. The total number of iterations is $2 \times 10^5$. The number of middle nodes are set larger than in section \ref{subsub_section_nodelay_small} because larger number of nodes are necessary to generate time delay. In this section, the target pattern is $\mathbb{T}=A_{\tau_d}$ and we consider various delay intervals $\tau_d$. In all cases the time windows for the response is $\tau_r = 30$.  
	
\begin{figure}[!ht] 
		\begin{center}
			\includegraphics[width=1.0\columnwidth, clip]{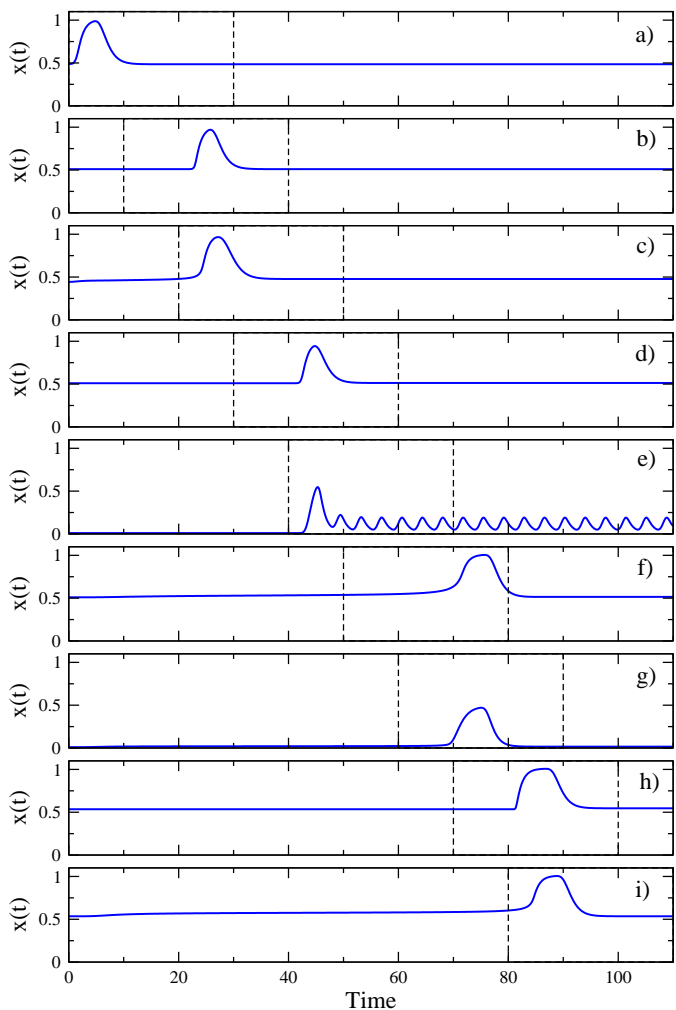}
			\caption{Examples of output signals constructed by requiring different delays $\tau_d$. The response windows $\tau_r=30$ are shown between dashed lines. The network errors are: $\epsilon_a = 0.001$, $\epsilon_b=0.021$, $\epsilon_c=0.011$, $\epsilon_d=0.036$, $\epsilon_e=0.045$, $\epsilon_f=0.038$, $\epsilon_g=0.033$, $\epsilon_h=0.018$ and $\epsilon_i=0.094$.}
			\label{fig_fig11} 
		\end{center}
	\end{figure}

	We run several realizations and we find networks with the predefined target pattern. Figure \ref{fig_fig11} presents output adaptive signals for several  networks, each of them with a different required delay $\tau_d$. The obtained networks are stored in the supplementary material.  In general we find that responses with shorter delays are easier to obtain than responses with longer delays. In Fig. \ref{fig_fig11} the external signal activates the receptors at $t=0$. Despite the differences in the delays of the signals, all networks have the same number of nodes and connections.

	We note that there are many different pulse shapes and pulse durations. The behavior of these networks is quite interesting since the characteristic time of the dynamics of a node is given by the constant $\gamma$ which has a unit value in this model, however, very long delays can be constructed. Thus, the delays with several order of magnitude larger than the characteristic time $\gamma$ are realized with $M=5$ middle nodes and $L=15$ links.

	\subsubsection{One receptor and three output nodes}

	In this example we construct a network with one receptor and three output nodes and delayed responses. The aim is to activate these three output nodes with different time delays after the activation of the receptor. The target matrix is the following one:

	\begin{equation}
		\mathbb{T} = \left[
		\begin{array}{c}
			A_{0} \\
			A_{30} \\
			A_{60}
		\end{array}
		\right]
		\label{matrix_target_1x3}
	\end{equation}

	\begin{figure}[!ht] 
		\begin{center}
			\includegraphics[width=1.0\columnwidth, clip]{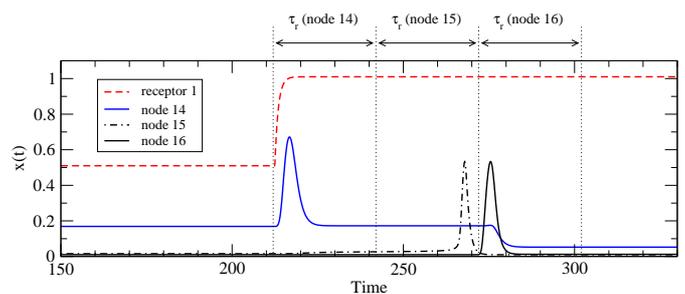}
			\caption{Output signals of a network with one input receptor and three output nodes (from Fig. \ref{fig_fig13}). The response interval $\tau_r=30$ is shown between vertical dashed lines. }
			\label{fig_fig12} 
		\end{center}
	\end{figure}

	\noindent
	The optimization has $2 \times 10^5$ iterations and the temperature parameter $\log(\sigma)=-2$. The response windows has $\tau_r=30$. We employ a rewiring scheme of evolutionary mutation. The network has $M=12$ middle nodes and $L = 36$ connections. 

\begin{figure}[!ht] 
	\begin{center}
		\includegraphics[width=1.0\columnwidth, clip]{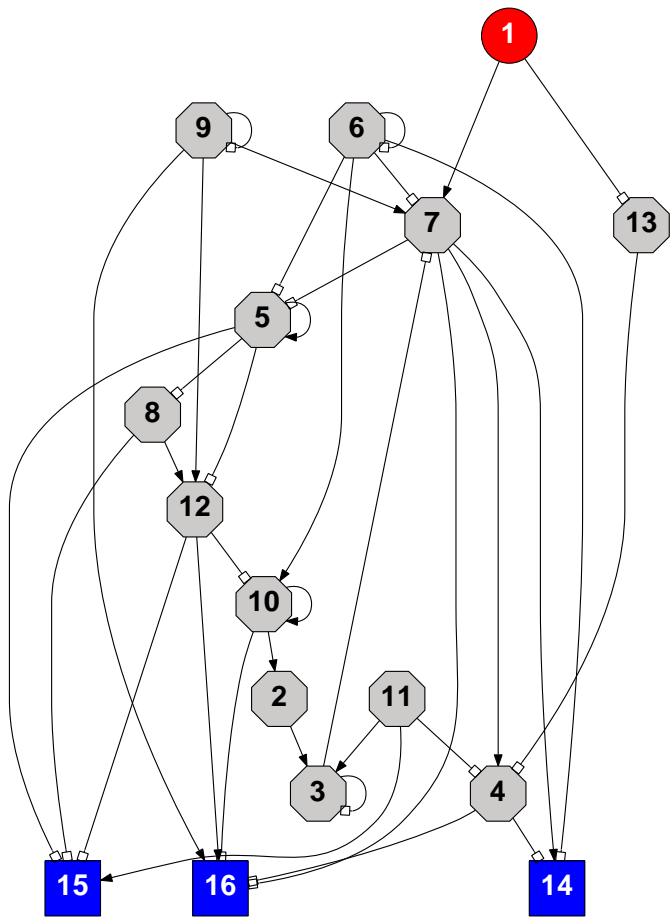}
		\caption{Network constructed to reproduce the target pattern of the matrix \ref{matrix_target_1x3}. Its output signals are shown in Fig. \ref{fig_fig12}. The network error is $\epsilon = 0.012$. }
		\label{fig_fig13}
	\end{center}
	\end{figure}

	Figure \ref{fig_fig12} shows the output signal of the network (Fig. \ref{fig_fig13}) constructed to produce the required target pattern (matrix \ref{matrix_target_1x3}). The dashed red curve shows the expression of the receptor. We observe that the three pulses are located inside their required time windows of response.  

	In this particular case of target pattern, it is possible to construct the required responses by merging three networks with one receptor and one output node each. For example taking the networks constructed in sec. \ref{sec_1_1_delays} with the proper delays. Merging these networks by their receptor nodes, we have a new network with one receptor and three output nodes connected through each independent route. That network should have a quite different architecture from the one we have found here. Figure \ref{fig_fig13} shows how a middle node can be involved in several pathways responsive of output signals with different delays.

	\subsubsection{Three receptors and one output node}

	We consider a network with three input receptors and one output node with delayed responses. In this case the output node must generate three different adaptive responses with different time delays $\tau_d$ as a function of the activated receptor. The target matrix is the following:
	
	\begin{equation}
		\mathbb{T} = \left[
		\begin{array}{c c c}
			A_{0} & A_{30} & A_{60}
		\end{array}
		\right]
		\label{matrix_target_3x1}
	\end{equation}

	\begin{figure}[!ht] 
		\begin{center}
			\includegraphics[width=1.0\columnwidth, clip]{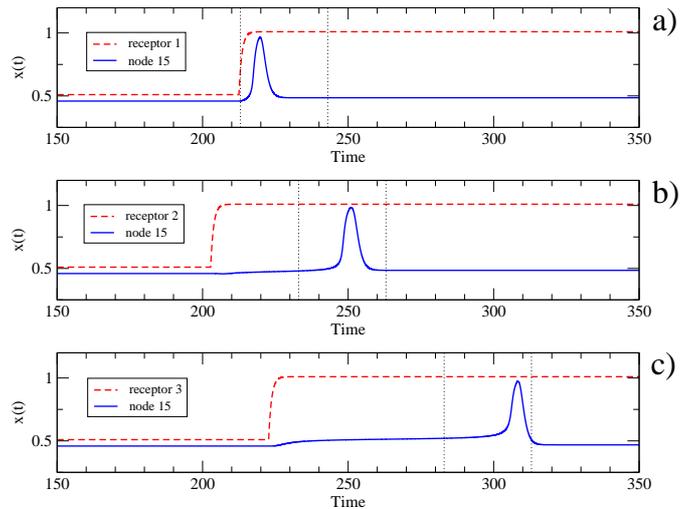}
			\caption{Output signals of the network from Fig. \ref{fig_fig15}. The responses intervals $\tau_r=30$ are shown between vertical dashed lines.}
			\label{fig_fig14} 
		\end{center}
	\end{figure}

\begin{figure}[!ht] 
		\begin{center}
			\includegraphics[width=1.0\columnwidth, clip]{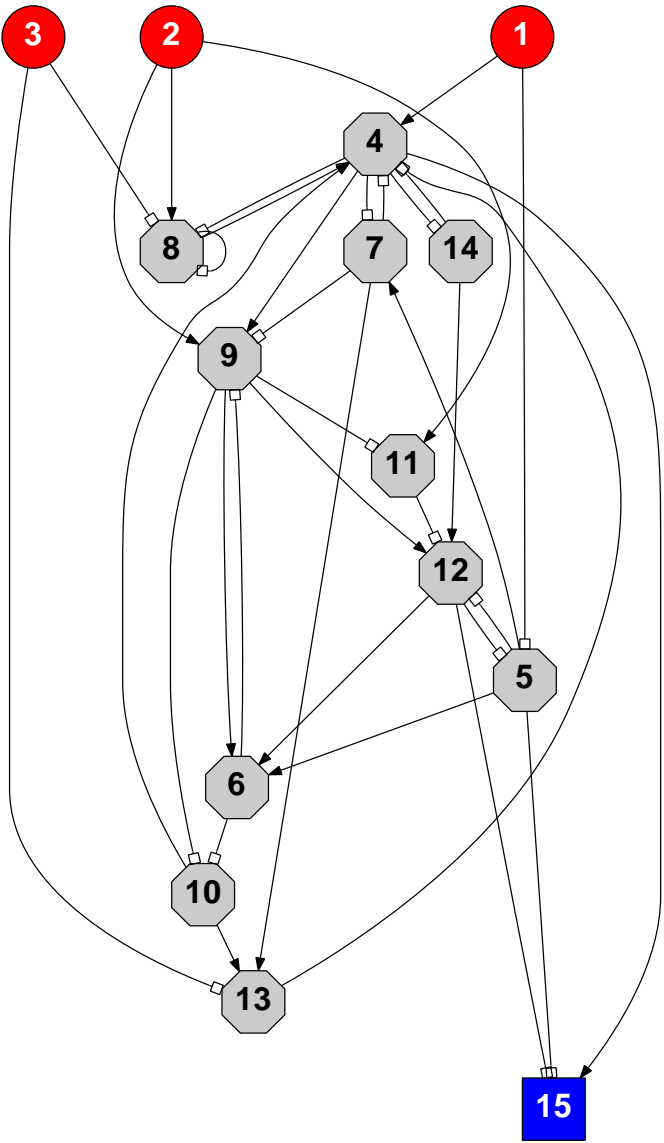}
			\caption{Constructed network able to reproduce the target pattern of matrix \ref{matrix_target_3x1}. Its output signals are shown in Fig. \ref{fig_fig14}.The network error is $\epsilon =0.017$.}
			\label{fig_fig15} 
		\end{center}
	\end{figure}

	For the optimization we use  $M=12$ middle nodes and $36$ connections, $\tau_r=30$,  $2 \times 10^5$ iterations and $\log(\sigma)=-2$. We employ the rewiring scheme of evolutionary mutation. In Fig. \ref{fig_fig14} we show the output signal of the solution network in Fig. \ref{fig_fig15}. We can observe that the three pulses are generated according to the target pattern. Note that the final networks has $M=11$ middle nodes, since that at the end of the simulation we find an isolated node and it was deleted from the network.
	
	Contrary to the previous case (section 3.2.2), here the solution cannot be constructed by merging networks. In effect, we can take three networks designed for the proper delayed adaptive responses with one receptor and one output node, and merge them by their output nodes. However, the new output signal is in general different from the superposition of individual network signals since the non linearities of the dynamical system.

	\subsubsection{Three receptors and three output nodes with delayed responses}

	In this last example, we construct a relatively big network with three input receptors, three output nodes, and, different delayed responses. The target pattern is the following one:

	\begin{equation}
		\mathbb{T} = \left[
		\begin{array}{c c c}
			R_{0} 	& 0	 & A_{60} \\
			0 	& A_{30} & R_{60} \\
			A_{0} 	& R_{30} & 0 \\
		\end{array}
		\right]
		\label{matrix_target_3x3_d}
	\end{equation}

	\noindent
	Note that this target pattern is similar to the one of the previous example in sec. \ref{subsub_section_3x3_1}, with the matrix \ref{matrix_activation}. However, now we require that the responses as a consequece of the activation of a receptor have the same delay, and these delays are different for each receptor. For the construction we use $M=25$ middle nodes and $L=80$ connections. The time windows for the response is $\tau_r=30$. The optimization was performed with the rewiring scheme of mutation and we use as temperature parameter $\log(\sigma)=-2$. The total number of iterations is $2 \times 10^5$.

	\begin{figure}[!ht] 
		\begin{center}
			\includegraphics[width=1.0\columnwidth, clip]{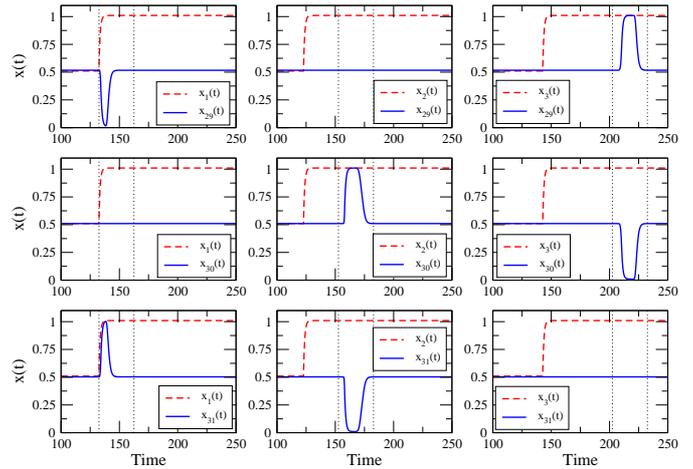}
			\caption{Output pattern of a network constructed with $M=25$ middle nodes and $L=80$ links (Fig. \ref{fig_fig17}). The response intervals ($\tau_r=30$) are shown between vertical dashed lines.}
			\label{fig_fig16} 
		\end{center}
	\end{figure}
	
\begin{figure}[!ht] 
		\begin{center}
			\includegraphics[width=1.0\columnwidth, clip]{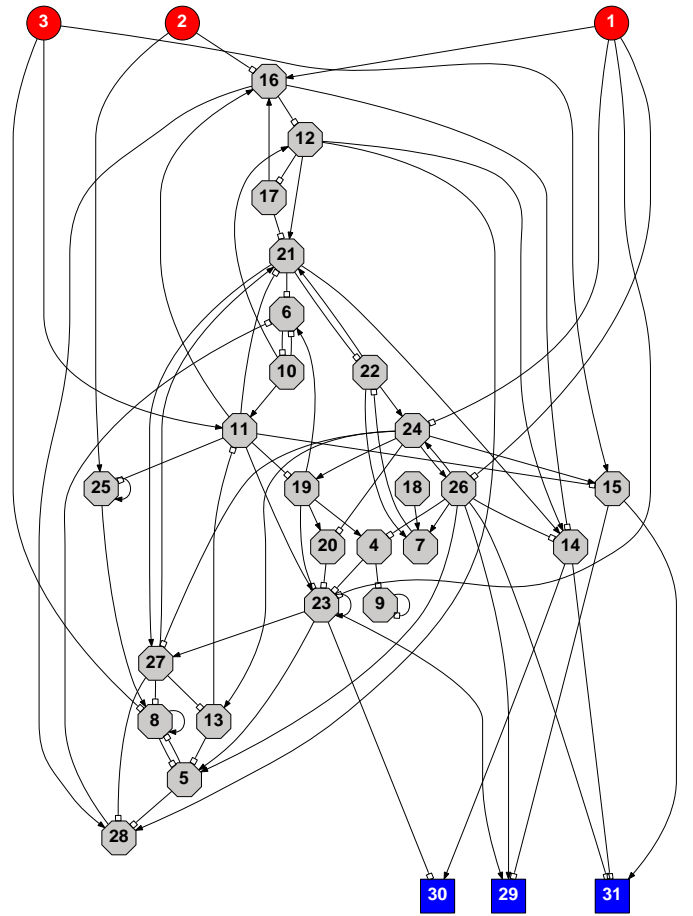}
			\caption{Network with the output shown in Fig. \ref{fig_fig16}. The network error is $\epsilon = 0.001$.}
			\label{fig_fig17} 
		\end{center}
	\end{figure}

	Figure \ref{fig_fig16} shows the ouput signals of the constructed network, and the network is shown in Fig. \ref{fig_fig17}. We observe that each receptor generates three different responses on the output nodes with the same delay. Red dashed curves indicates the epression level of the receptors and the blue continue curves the adaptive responses on the output nodes. In this network, all the responses are correctly performed and the network presents a small error. The example shows that relatively big networks with delayed responses can be constructed by our method.
	
	\section{Discussions and conclusions}
	\label{sec_conclusion}
	
	In this work we have presented a method to construct networks with several receptors and output nodes able to generate adaptive delayed responses. We show several examples of network constructions and statistical analysis of ensembles of functional networks.
	
	The annealing algorithm used in this work is a powerful tool in order to construct systems with a required target. Although we can find solutions for a proposed problem, we cannot know in advance if a given number of nodes and links are enough in order to solve the required problem, and even to guarantee the existence of a solution. For this reason, the examples we presented are the networks construed with the smaller number of nodes and links as we could that can reproduce the target pattern. 
	
	Different from previous works \cite{Trusina,book_zhang,Inoue}, our networks can generate several responses as a function of the activated receptor. One important characteristic of these networks is that a middle node is in general involved in the activation of many output nodes, and it can be used as processing unit for many receptors. This characteristic implies complex structures for the constructed networks.   
	
	The architecture by layers and the connection rules between nodes allow us to define specific functions for the nodes (receptors, middle nodes, output nodes). Thus, this structure helps to study the functionality of big networks. On the other hand, the layered-like structure differs from the classical architectures  \cite{Trusina,book_zhang,Inoue} of networks with adaptive response with three nodes. In effect, we cannot obtain adaptive responses with only one middle node. However, our algorithm can apply to any architecture where a cost function is well defined. 
	
	In this work the main objective is to construct functional networks with delayed responses. We do not analyze the mechanism involved in the delay generation. However, we can mention that the generation of delayed responses by chaining small subnetworks which already present delays, is known \cite{delayed_response}. In our results, this kind of solution is not found, as we can observe from the network structures. Our hypothesis is that the delays are related to a slow dynamics generated by setting the system close to a bifurcation point as a function of the activation signal. Thus, after the transient the system settle down on a stable fixed point. When the input signal is turned on, the fixed point disappear, the remained vector field has small velocity, and, the system moves slowly to the new attractor. This situation is well known for saddle-node bifurcation and its bottle neck effects on the dynamics. This hypothesis is current work in progress of the authors.
	
	We thank to Tatsuo Shibata for his comments and observations about this manuscript. P.K. acknowledges financial support from SeCTyP-UNCuyo (project M009 2013-2015) and from CONICET (PIP 11220150100013), Argentina and M.I. acknowledges financial support from JSPS Research Fellowships for Young Scientists. Both authors contributed equally to the paper.

\end{document}